# Revealing subterahertz atomic vibrations in quantum paraelectrics by surface-sensitive spintronic terahertz spectroscopy

Zhaodong Chu[1]†, Junyi Yang[1]†, Yan Li[1], Kyle Hwangbo[2], Jianguo Wen[3], Ashley R. Bielinski[1], Qi Zhang[4,5], Alex B. F. Martinson[1], Stephan Hruszkewycz[1], Dillon D. Fong[1], Xiaodong Xu[2], Michael R. Norman[1]*, Anand Bhattacharya[1]*, Haidan Wen[1,4]*

[1]Materials Science Division, Argonne National Laboratory, Lemont, IL 60439, USA

[2]Department of Physics, University of Washington, Seattle, Washington 98195, USA

[3]Center for Nanoscale Materials, Argonne National Laboratory, Lemont, IL 60439, USA

[4]Advanced Photon Source, Argonne National Laboratory, Lemont, IL 60439, USA

[5]Present address: Department of Physics, Nanjing University, China

†These authors contributed equally to this work.

E-mail: norman@anl.gov; anand@anl.gov; wen@anl.gov

**Abstract**

Understanding surface collective dynamics in quantum materials is crucial for advancing quantum technologies. For example, surface phonon modes in quantum paraelectrics are thought to play an essential role in facilitating interfacial superconductivity. However, detecting these modes, especially below 1 terahertz (THz), is challenging due to limited sampling volumes and the need for high spectroscopic resolution. Here, we report surface soft transverse optical (TO1) phonon dynamics in $KTaO_3$ and $SrTiO_3$ by developing surface-sensitive spintronic THz spectroscopy that can sense the collective modes only a few nanometers deep from the surface. In $KTaO_3$, the TO1 mode softens and sharpens with decreasing temperature, leveling off at 0.7 THz. In contrast, this mode in $SrTiO_3$ broadens significantly below the quantum paraelectric crossover and coincides with the hardening of a sub-meV phonon mode related to the antiferrodistortive transition. These observations that deviate from their bulk properties may have implications for interfacial superconductivity and ferroelectricity. The developed technique opens opportunities for sensing low-energy surface excitations.

**Teaser:** Ultrafast terahertz emission captures surface soft phonon modes that deviate from their bulk properties in quantum paraelectrics.

**MAIN TEXT**

**Introduction**

Surface phonons - collective vibrations of atoms at an interface or surface that span only a few nm along the depth direction - are pivotal to surface dynamics and properties. Their unique characteristics can deviate significantly from the bulk and have been found to underlie novel nanoscale phenomena such as surface phonon polaritons[1], interfacial superconductivity[2-5], surface catalysis[6], and interfacial thermal transport[7]. Understanding and controlling these collective dynamics at the surface are essential for applications. For instance, in quantum paraelectrics such as $KTaO_3$ (KTO) and $SrTiO_3$ (STO), recent studies[3-5] suggest that the observed interfacial superconductivity is driven by the coupling of the electrons to the low energy surface transverse optical (TO1) phonon. Variations in the superconducting transition temperature with interface orientation in these quantum paraelectrics were proposed to result from the distinct surface TO1 mode dynamics unique to each material[4,5]. However, direct experimental evidence is challenging to obtain due to the lack of experimental techniques capable of probing low-energy surface phonon modes (< 1THz) with the required spectroscopic resolution and limited sampling volumes.

Conventional tools such as Raman[8], hyper Raman[9], and terahertz (THz) spectroscopy[10] rely on far-field light-matter interactions with the optical penetration depth typically on the order of tens to thousands of nanometers and are thus not sensitive to surface modes. The advances in tip-enhanced photonic techniques, such as THz scanning near-field optical microscope[11,12], provide near-field light-matter interactions. However, due to the long wavelength of the THz radiation (~300 μm), the minimum probing depth can only approach tens of nm. Similar challenges remain for tip-enhanced Raman spectroscopy[13], where it is difficult to separate the bulk contribution from the surface one. Planar tunneling into conducting STO, though sensitive well below the meV level, has not been able to detect the coupling of electrons to the low-energy transverse optical phonons[14,15]. Other scattering spectroscopies such as laser-induced scanning tunneling microscopy[16], surface-enhanced Raman scattering[17], electron energy-loss spectroscopy[18,19] and helium atom scattering[20] can achieve the needed surface sensitivity, but their energy resolution is limited to tens of meV. On the other hand, neutron scattering[21] can achieve sub-meV energy resolution, but it falls short in measuring surface modes due to the substantial sample volume required for measurements. In this context, sensitive probes of low-energy collective dynamics that are confined within a few nanometers from a bulk crystal surface with meV energy sensitivity are urgently needed.

Here, we introduce a surface-sensitive spintronic THz spectroscopy (SSTS) where both the cavity-enhanced near-field THz-matter interaction and ultrafast coupling between the interfacial transient current and the sample are confined to the nanoscale in the vicinity of the sample surface for detecting ultralow-energy surface modes. The probing depth of SSTS in KTO and STO was found to be ~ 5 nanometers. Our approach differs from traditional THz surface spectroscopy[22-25], where the absorption depth of the far-field optical pump light defines a probe depth typically on the order of tens to hundreds of nm. Although this depth can be regarded as a surface probe in a certain context[25], it is significantly deeper than the few nanometers near an interface where exotic quantum processes occur such as interfacial superconductivity in quantum paraelectrics[2-5]. Using SSTS, we uncovered the distinct dynamics of surface meV phonons that have thus far been elusive in two representative quantum paraelectric crystals, KTO and STO. Our temperature-dependent phonon spectra revealed an unexpected broadening of the TO1 mode at the STO surface, in contrast with the sharp TO1 mode found in bulk STO as well as at the KTO surface. These observations potentially shed light on the origin of differing interfacial superconducting transitions among quantum paraelectrics, as the surface TO1 mode is thought to play an important role in the Cooper pairing process. Our work showcases SSTS's capability to provide unprecedented insights into surface and interfacial dynamics in quantum materials.

**Results**
*Surface-Sensitive Spintronic THz spectroscopy (SSTS)*
The design of SSTS is based on THz spintronic emission via spin-to-charge current conversion, where a transient current (i.e., the THz source) is confined within a nm-thin metallic layer or interface. Previous THz spintronics studies based on metal/metal[26-32] and metal/insulator[33-38] heterostructures primarily focused on optimizing efficiency and bandwidth as an intense THz source and exploring various spin-to-charge conversion mechanisms. This work illustrates that SSTS based on dielectric-ferromagnetic metal heterostructures can probe the collective dynamics confined within a few nanometers of the underlying substrate's surface. As exemplified by a Py/KTO (111) heterostructure shown in Fig. 1A, a 3-nm-thick, in-plane magnetized ferromagnet (FM) layer, permalloy (Py: $Ni_{0.8}Fe_{0.2}$), was deposited on the single crystal sample, with atomically sharp interfaces (Methods and Fig. S1). An 800 nm, 100 fs laser pulse is passed through the transparent sample, exciting the Py layer (Methods). The resulting spin super-diffusion current $J_0(t)$ (the black arrow in Fig. 1A) is converted into an in-plane, spin-polarized charge current $J_c(t)$ via spin-to-charge conversion at the interface[36-39] (the downward-pointing blue arrow in Fig. 1A). This interfacial transient current $J_c(t)$ serves as the source of the THz radiation, and the resulting far-field radiation $S(t)$ is detected by free-space, electro-optic sampling (Methods and Fig. S2). To verify the spintronic origin of the THz generation via spin-to-charge conversion, rather than the photo-Dember effect[22,23] or the surface surge currents[24] reported in semiconductors, we measured the expected dependence of the THz signal on the magnetization direction in the Py/KTO and Py/STO samples, and we also found no THz emission from either bare KTO (or STO) crystals or non-magnetic metal (NM)/STO heterostructures under the same laser excitation conditions (Fig. S3). Furthermore, we confirmed that the contribution of ultrafast demagnetization-induced magnetic dipole radiation in our samples is negligible (Fig. S4).

Fig. 1B displays the measured THz waveform of the Py/KTO (111) sample and its Fourier spectrum at the temperature T = 10 K. A prominent spectral dip at the TO1 phonon mode frequency of the sample is evident in the emitted THz spectrum, in contrast to the broadband THz emission in a sapphire reference sample which is featureless in the measured THz window (Fig. S5). We demonstrated that the TO1 spectral dip is not due to the THz absorption by the KTO substrate by carrying out far-field THz reflection measurements (detailed in Fig. S6). Fig. 1C shows the reflected THz spectrum from a KTO (111) substrate at T = 10 K, where no spectral dip is observed.

Moreover, we found that spatially separating the THz source ($J_c(t)$) from the sample by adding THz featureless spacers (Fig. S7), such as 100-nm-thick $SiO_2$ or a few micron-thick air gap, between the sample and the Py layer, prevents the detection of the TO1 spectral dip in the emitted THz radiation, even though the THz field in the Py layer can still pass through the spacer layer to probe KTO and the reflected THz field was recorded. These results confirm that the absorption of far-field THz radiation is not the mechanism of SSTS. Our theoretical modeling of SSTS suggests that the detection mechanism has two contributors. One is cavity-enhanced near-field THz-matter interaction at the interface, where the 3 nm Py layer acts as an ultra-thin Fabry-Pérot cavity, allowing the near-field THz to directly sense the cavity impedance and the dielectric response of the interfaces. The other is the polarization modulation at the KTO surface, induced by the interfacial current $J_c(t)$ due to hot electron-TO1 phonon coupling, which interferes destructively with $J_c(t)$. As detailed in the "Modeling SSTS" section later in the main text, the two contributions enable the surface TO1 mode to be detected as a spectral dip in the emitted THz field.

To experimentally quantify the probe depth of the surface TO1 mode in KTO and establish the interface sensitivity of SSTS, we carried out experiments in which the thickness of a buffer layer between the Py and the KTO substrate was varied. We chose $BaTiO_3$ (BTO), another member of the perovskite oxide family, as the buffer layer, which has a similar electrical band gap and high dielectric constant as the substrate but does not host any modes within our THz detection window[40]. The BTO-buffer layer was deposited by pulsed laser deposition (see Methods) to achieve atomically sharp interfaces with nanometer thickness (Fig. S8) so that the leakage current was eliminated. Fig. 1D presents the emitted THz spectrum $S(\omega)$ from Py/BTO-buffer/KTO (111) samples at $T = 10$ K as a function of the buffer layer thickness. The spectral dip positions vary as the buffer layer thickness increases from 1 nm to 4.4 nm, suggesting that the spectral dip does not arise from a bulk phonon mode as the substrate is the same for these samples. When the thickness of the buffer layer is increased beyond 5.7 nm, no spectral dip is observed in the detection window. This result demonstrates that a BTO-buffer layer with a thickness larger than 5.7 nm effectively prevents the detection of the KTO surface TO1 phonon. We quantified the critical BTO-buffer layer thickness ($d_c$) to be ~ 5 nm as the probe depth for detecting the KTO TO1 mode, as shown in Fig. 1E. Such a probe depth matches the width of the two-dimensional electron gas (2DEG) formed at KTO or STO interfaces/surfaces[41]. Therefore, SSTS is a powerful tool for revealing the dynamics of the surface TO1 phonons involved in the unconventional superconductivity at quantum paraelectric interfaces. In addition, SSTS does not require a strong nonlinear optical response at the interface as needed for the differential frequency generation process[42].

### *Surface phonon dynamics of KTaO$_3$ and SrTiO$_3$*

By applying SSTS, we uncovered the dynamics of surface meV phonon modes for two representative quantum paraelectrics, KTO and STO. Although the temperature-dependent evolution of the THz waveforms in both samples exhibited similarities at high temperatures (as indicated by the green curves in the time trace at 200 K in Figs. 2A and 2B), distinct differences emerged in the low-temperature regime, as highlighted by the time traces at 10 K (purple curves in Figs. 2A and 2B). The evolution of the THz spectra was primarily driven by the temperature-dependent surface TO1 phonon dynamics (Figs. 2C and 2D). At temperatures above 200 K, the spectral dips were less pronounced due to the limited THz detection window. At low temperatures, distinct spectral features from surface modes were observed. For KTO (Figs. 2C and 2E), the dip corresponding to the TO1 mode sharpened as the temperature decreased, consistent with reduced thermal fluctuations. Moreover, the softening of the mode ceased below ~15 K, leveling off at a mode frequency of 0.7 ± 0.02 THz (the error bar of the mode frequency is determined in Fig. S9). This observation was reliably reproduced in measurements on two other Py/KTO (111) samples prepared in the same way, as shown in Fig. 1B and Fig. S10. The THz spectra of the Py/STO (111) sample (Figs. 2D and 2F) exhibited characteristics at low temperatures that differ from what was observed in Py/KTO (111). In the STO sample, the surface TO1 mode dip was shallower and broader at temperatures below ~ 60 K compared to KTO and eventually became featureless below ~ 35 K, the crossover temperature between paraelectric and quantum paraelectric behavior[43,44]. Moreover, a sub-meV mode, highlighted by the red dashed curve in Fig. 2D and red arrows in Fig.

2F, emerged below ~100 K and then hardened as the temperature was further lowered. The spectroscopic dip for this lower-energy mode, similar to the TO1 mode, broadened as the temperature decreased and was not discernable at temperatures below approximately 35 K. The emergence of this surface sub-meV (< 0.25 THz) mode below 100 K is likely associated with the antiferrodistortive transition from the cubic to the tetragonal phase[45-47]. In bulk STO, the antiferrodistortive transition occurs at 105 K, related to phonon softening at the *R* point of the Brillouin zone. Below this transition, the unit cell doubles, allowing the *R* point phonon to manifest at the *Γ* point. Although this phonon has even parity, inversion symmetry breaking at the interface can render it IR active, leading us to conjecture that this is the origin of the emergent sub-meV mode. As expected, this phonon hardens below the antiferrodistortive transition. In bulk STO, two phonons appear below 105 K as the triply degenerate *R* phonon splits into a doubly degenerate *E* mode and a singly degenerate *A* mode due to the symmetry reduction from cubic to tetragonal. The lower-energy *E* mode at low temperatures is approximately 0.45 THz in crystals[46] and 1.2 THz in polycrystalline ceramics[47], both significantly higher than the sub-meV mode we observed. This suggests that the *R*-point phonon may be softer at the surface than in the bulk, or that the splitting of the *A* and *E* modes is more pronounced at the surface or interface (note that the A mode is not IR active[47]). Additionally, the two broad surface phonons (i.e., the sub-meV mode and the TO1 mode) merge at low temperatures (Figs. 2D and 2F), suggesting an interaction between the two modes that is not evident in STO bulk samples. Thus, our observations show a strong deviation between bulk STO and the surface.

To further compare the TO1 mode at the surface and in the bulk of both KTO and STO, we conducted the following analysis. Fig. 3A illustrates the frequencies of the surface and bulk TO1 modes in KTO as a function of temperature. Notably, above 30 K, the mode frequencies of the surface and bulk were approximately the same, following a temperature dependence that is consistent with the Curie-Weiss law (i.e, $\omega_{TO1} \propto \sqrt{T - T_c}$). Below 30 K, the softening of both the surface and bulk modes was gradually suppressed, deviating from the Curie-Weiss law at the paraelectric to quantum paraelectric crossover temperature. The surface TO1 frequency was found to be ~ 0.1 THz higher than that of the bulk KTO mode reported in the literature[45,48], likely due to the presence of a static internal electric field at the Py/KTO interface[49].

A comparison of the TO1 mode at the STO surface and the bulk is shown in Figs. 3B and 3C. Above 60 K, the TO1 mode frequency for both surface and bulk aligned with the Curie-Weiss law, with a fitted transition temperature $T_c$ of 37 K. Below 60 K, the softening deviated from the Curie-Weiss law, as expected. However, unlike KTO, the STO surface phonon appears to be softer than the bulk one. Similar broadening of the THz spectroscopic dip from the surface TO1 phonon was also observed for a Py/STO (001) sample in the same temperature range (Fig. S10). The broadening of the surface phonon linewidth observed by SSTS at low temperatures has not been previously reported in STO crystals[45,47,50] or films[49,51]. Prior hyper-Raman spectroscopy (HRS) measurements[50] (Fig. 3C) showed that the bulk TO1 mode linewidth sharpens as the temperature decreases, opposite to what we observed at the surface with SSTS. This sharpening of the bulk linewidth was corroborated for the $A_{1g}$ and $E_g$ phonons measured by Raman spectroscopy on the same Py/STO (001) sample in this study (Fig. S11).

The significant broadening of the surface TO1 mode below the quantum paraelectric crossover indicates strong fluctuations of the TO1 modes at the STO surface. Such fluctuations may affect how the surface TO1 mode mediates Cooper pairing and therefore the interfacial superconducting transition temperatures. For example, the transition temperature for superconductivity of the 2DEG at the STO (111) interface is in the range of hundreds of mK, an order of magnitude lower than that for the 2DEG at the KTO (111) interface[3,4]. Similarly, the surface TO1 spectral dips of a KTO (001) 2DEG sample, where the 2DEG remains normal down to tens of mK[3,4], are also much broader than those of the KTO (111) 2DEG (Fig. S12). The fluctuation being more pronounced in STO than in KTO could be due to STO's closer proximity to the ferroelectric quantum critical point[52], which makes the behavior of STO more dependent on its surface or interface reconstruction and imperfections[53-55], leading to spectral broadening of the surface TO1 mode. Although these results alone would not explain why doped bulk STO and STO (111) interfaces exhibit similar superconducting transition temperatures[56,57], our findings offer valuable insights into the interfacial superconductivity of quantum paraelectrics. Additionally, they underscore the exceptional capabilities of SSTS in revealing low-energy interfacial structural dynamics.

*Modeling SSTS*

As shown in our experiments (Fig. 1C, 1D, Figs. S6 and S7), the underlying mechanism of SSTS in detecting surface modes is not due to the absorption of the THz waves because the reflected THz wave at the far field does not encode the phonon information. To understand the underlying detection mechanism of SSTS, we model the cavity-enhanced near-field THz-matter interactions and electron-phonon couplings as detailed below.

The purpose of depositing a 3-nm-thick Py layer on the sample surface is twofold: it serves as the source for spintronic THz generation as well as provides a Fabry-Pérot cavity. This cavity confines the generated THz field, thereby enhancing near-field THz-matter interactions at the interface. The generated near-field THz $E_{near}(\omega)$ inside the cavity can be approximately expressed as a generalized Ohm's law[26,27]:

$$E_{near}(\omega) = Z(\omega)J_c(\omega) = \frac{Z_0}{n_{air}(\omega)+n_{samp}(\omega)+Z_0 G}J_c(\omega). \qquad (1)$$

Here, $E_{near}(\omega)$ and $J_c(\omega)$ are the Fourier transforms of $E(t)$ and $J_c(t)$, respectively. $Z(\omega)$ is the impedance of the cavity, $Z_0$ is the vacuum impedance, and G is the conductance of the Py film. $n_{air}$ and $n_{samp}$ are the refractive indices of air and the sample (i.e., KTO or STO), respectively. The measured THz spectrum $S(\omega)$ is the far-field radiation which is convolved with the propagation of $E_{near}(\omega)$ from the Py to the detector and the detector response function $H(\omega)$, i.e., $S(\omega) = E_{near}(\omega) \cdot H(\omega)$. From Eq. (1), there are two contributions to the detection of the surface modes such as the TO1 mode. As depicted in Fig. 4A, one is the impedance of the cavity, $Z(\omega)$, which characterizes the near-field THz-matter interaction at the interface via its dielectric response as the spectral information of the surface TO1 mode is encoded in the dielectric function. The other is the transient current source term, $J_c(\omega)$, which can be modulated by the surface polarization, $dP(t)/dt$, induced by hot electron - TO1 phonon interactions. We quantitatively examine the contribution of the two terms separately, as described below.

We first focus on the contribution of the impedance term $Z(\omega)$ by making the approximation that $J_c(\omega)$ has no frequency structure due to the surface modes and thus assume that the ratio of $J_c(\omega)$ for the sample (Py/KTO or Py/STO) to that of the reference (Py/sapphire) is a constant, i.e. $\delta_J = \frac{J_{c,samp}(\omega)}{J_{c,ref}(\omega)} = C$. By dividing the experimental THz spectrum for the sample ($S_{exp,samp}(\omega)$) and the reference ($S_{exp,ref}(\omega)$), one obtains:

$$\frac{S_{exp,samp}(\omega)}{S_{exp,ref}(\omega)} = \frac{J_{c,samp}(\omega)\,[1+n_{ref}+Z_0 G]}{J_{c,ref}(\omega)\,[1+n_{samp}(\omega)+Z_0 G]} = C\frac{1+n_{ref}+Z_0 G}{1+n_{samp}(\omega)+Z_0 G}. \quad (2)$$

This division has the additional benefit of removing $H(\omega)$. Based on Eq. (2), the THz signal from KTO or STO is expected to be suppressed relative to the reference because the refraction index of the KTO or STO ($n_{samp}$) is much larger than that of sapphire ($n_{ref}$=3.31) and $Z_0 G$ (estimated to be 1.7 -- see Fig. S13 caption), in agreement with our data (Fig. S13). Assuming Eq. (2) is valid, the index of refraction of the sample surface can be determined solely from the measured THz spectra (Fig. S13). We can also use Eq. (2) to estimate the THz spectrum at 6 K for KTO from the contribution of $Z(\omega)$ alone from $S_{Z,KTO}(\omega) = C\frac{S_{exp,ref}(\omega)\,[1+n_{ref}+Z_0 G]}{[1+n_{KTO}(\omega)+Z_0 G]}$ where $S_{exp,ref}(\omega)$ is the experimental data of Py/sapphire at 6 K. $n_{KTO}$ is modeled using the Lyddane-Sachs-Teller relation[58], $n_{KTO}(\omega)^2 = \varepsilon^*(\omega) = \varepsilon(\infty)\prod_i^n \frac{\omega_{LOi}^2 - \omega^2 + i\omega\gamma_{LOi}}{\omega_{TOi}^2 - \omega^2 + i\omega\gamma_{TOi}}$, where the frequencies of the various longitudinal ($\omega_{LOi}$) and transverse ($\omega_{TOi}$) optical phonon modes are from IR measurements[58]. To make a more robust comparison to our experimental data $S_{exp,KTO}(\omega)$ at 6 K, we adjusted the TO1 mode frequency from the IR value of 0.6 THz to 0.71 THz to match the location of the spectral dip, and we fixed C (1.62) by matching to the data at the upper range of our THz window.

The probing depth of the cavity impedance $Z(\omega)$ based on the dielectric response is determined by the depth of the near-field enhanced region close to the cavity. This depth can be as short as a few nm in materials with a large dielectric constant, as we experimentally demonstrated using a BTO buffer layer (Fig. 1D). To assess this depth for low dielectric materials, we replaced the BTO buffer layer with $Al_2O_3$ and found that the probe depth of $Z(\omega)$ in $Al_2O_3$ extends from tens of nanometers to a hundred nanometers (Fig. S14). Therefore, the strong resonance of the TO1 mode in our THz window that gives rise to a large dielectric response is important to shorten the probe depth of the $Z(\omega)$ term.

Fig. 4B shows the comparison of spectral profiles between $S_{Z,KTO}(\omega)$ and experimental results at 6 K, i.e., $S_{exp,KTO}(\omega)$ and $S_{exp,ref}(\omega)$. Although the spectrum of $S_{Z,KTO}(\omega)$ exhibits a THz intensity suppression very similar to $S_{exp,KTO}(\omega)$ when compared to $S_{exp,ref}(\omega)$, the TO1 dip strength is much weaker than that of $S_{exp,KTO}(\omega)$. This suggests that the current ratio $\delta_J = \frac{J_{c,samp}(\omega)}{J_{c,ref}(\omega)}$ is not a constant, and $J_{c,samp}(\omega)$ should have a frequency structure associated with the TO1 mode as well.

The registration of the surface TO1 mode in $J_c(\omega)$ can be understood as a modulation of the transient current[59] due to electron-lattice interactions at the interface, as illustrated in Fig. 4A. Both elastic and inelastic scattering at the Py/KTO interface can modify the charge current. Because of the limited frequency range of $J_c(\omega)$ set by the ultrafast current surge and relaxation in the Py layer as seen in the Py/sapphire reference, only those inelastic processes within this THz frequency range can be registered in the detected signal, as the spintronic THz emission process filters out other higher-frequency components (> 3 THz). For the present case of KTO or STO, the primary phonon involved in this process is the TO1 phonon. The conversion of the initial spin-polarized charge current into an in-plane charge current $J_c(\omega)$ plays an important role because the out-of-plane electron motion is converted to an in-plane one[36-39] allowing the electrons to couple to the transverse TO1 phonon[59].

To provide a quantitative description of this modulation of $J_c(\omega)$, we adopted an analytic model of the interaction of THz waves with a ferroelectric[60] that equally applies to the paraelectric case. We assume that the hot-electron-TO1 coupling leads to an ionic displacement field $D_{TO1}$, which produces a polarization modulation $dP/dt$ at the KTO interface with P proportional to the dielectric function, $\varepsilon(\omega)$. The induced contribution (i.e., the polarization modulation) has a phase lag of $\frac{\pi}{2} + \phi$ to the driving current $J_c(t)$, where $\frac{\pi}{2}$ comes from the time derivative and $\phi$ is the phase angle of the dielectric function ($\varepsilon(\omega) = |\varepsilon(\omega)|e^{i\phi}$). As $\omega$ sweeps through the TO1 mode frequency, this phase lag goes from $\frac{\pi}{2}$ to $\frac{3\pi}{2}$. At resonance, the phase lag is $\pi$, giving rise to destructive interference of this induced contribution with the driving current (Fig. 4C, inset). To illustrate this interference effect, if we model the time-dependent driving current as $J_c(t) \propto \cos(\omega t)$, then the induced contribution ($-dP/dt$) is $\propto c_{ind}|\varepsilon|\omega \sin(\omega t - \phi)$ with $c_{ind}$ a microscopic constant proportional to $d_P/c$ with $d_P$ the thickness of the polarization layer. Summing the two currents and Fourier transforming, this leads to an estimated current ratio $\delta'_J(\omega)$ in the frequency domain: $\delta'_J(\omega) = [(1 - c_{ind}|\varepsilon|\omega \sin(\phi))^2 + (c_{ind}|\varepsilon|\omega \cos(\phi))^2]^{1/2}$. Using the model dielectric function[58], Fig. 4C shows the profile of $\delta'_J(\omega)$ (the red curve) where the spectral dip arising from the surface TO1 mode is evident. Note that $c_{ind}$ is found to be small, so the large contribution of $-dP/dt$ to $\delta'_J(\omega)$ is due to the largeness of $\varepsilon$.

We now estimate the THz spectrum by considering the contribution from both the dielectric response as well as the current source term, $S_{J \cdot Z,KTO}(\omega) = \delta'_J(\omega) S_{Z,KTO}(\omega)$. As presented in Fig. 4B, $S_{J \cdot Z,KTO}(\omega)$ is in good agreement with the experimental data $S_{exp,KTO}(\omega)$ in both the spectral shape, intensity, and TO1 dip strength. At frequencies away from the TO1 mode, the contribution from $Z(\omega)$ predominantly suppresses the THz intensity compared to $S_{exp,ref}(\omega)$. However, at the TO1 mode frequency $\omega_{TO1}$, the intensity suppression relative to $S_{exp,ref}(\omega)$, i.e. $\frac{S_{exp,KTO}(\omega=\omega_{TO1})}{S_{exp,ref}(\omega=\omega_{TO1})}$, is about $10^{-2}$. This giant suppression is due in part to the contribution from the $Z(\omega)$, accounting for one order of magnitude $S_{Z,KTO}(\omega=\omega_{TO1})/S_{exp,ref}(\omega=\omega_{TO1})$, with the remaining order of magnitude attributed to the $J_c(\omega)$ term. Due to the needed interaction of the spin-polarized electrons with the sample, the probe depth of this mechanism, i.e. the $J_c(\omega)$ term, is limited to the tunneling depth of the electrons. In the case of KTO or STO[41], this is about a few nm which is consistent with the measurements (Fig. 1D).

The above discussion based on Eq (1) is focused on Py/dielectric samples without a buffer layer inserted. Eq (1) was derived under the assumption that the metal layer can be treated as a perturbation and that the electric field is constant within the metal layer and so does not take into account any depth (z) dependence of the dielectric response due to the formation of a Schottky barrier at the interface. This approximation becomes even more questionable with the additional presence of a buffer layer. Ignoring any z dependence, the inclusion of the buffer layer would have two effects: (i) it could introduce another source (current) term to the numerator of Eq. (1); (ii) it adds an imaginary constant to the denominator of Eq. (1) that is equal to [i($\omega$/c) $\varepsilon_B$ ·$d$] where $\varepsilon_B$ is the dielectric constant of the buffer layer and $d$ is the buffer thickness. This contribution from a buffer such as $SiO_2$ or $Al_2O_3$ with a thickness of 100 nm is of order $10^{-3}$ in our THz window and so negligible. Thus, the dielectric response at the KTO/STO interface should still show up in the presence of such a buffer layer, which contradicts our observations (Figs. S7 and S14). This would imply the necessity of a more complete treatment using the Poisson-Schroedinger equations to derive E(z), coupled with the wave equation for the THz radiation to calculate S($\omega$).

Our analysis based on Eq (1) and our experimental results uncover the detection mechanism of SSTS. Although previous studies[27,30] have considered the effect of the sample's dielectric response on the THz cavity radiation, this work unambiguously shown that both the cavity impedance term, $Z(\omega)$, and the source term, $J_c(\omega)$, contribute to detecting surface modes. Calculating the actual frequency profile of $J_c(\omega)$ as a result of the electron-lattice interaction would involve a detailed understanding of the interaction of the superdiffusive spin-polarized hot electrons in the Py with the phonons at the interface and presumably involves surface phonon dispersion and damping. Such complexities are beyond the present paper but would make an interesting topic for future study.

**Discussion**

We developed SSTS to enable direct experimental characterization of the sub-terahertz collective dynamics at surfaces that have not been achieved previously. This approach has revealed unique soft TO1 phonon dynamics at the surfaces of KTO and STO that deviate from the bulk, which is crucial for understanding interfacial superconductivity in quantum paraelectrics. In the context of quantum paraelectrics, our work provides an experimental platform to systematically study surface TO1 phonons that are hypothesized to be involved in the superconductivity of 2DEGs at their interfaces[3-5], for example in a superconductor heterostructure such as FM/AlO$_x$/KTO with an AlO$_x$ layer (~ a few nm thick) so that SSTS can effectively probe the interfacial TO1 phonon modes (Fig. S12). By gating this system, interfacial TO1 dynamics can be quantitatively measured as a function of the electric bias to explore the relationships between the TO1 phonon and interfacial superconductivity. Besides quantum paraelectrics, we envision our easy-to-use SSTS technique can be applied to a broad range of quantum materials for probing elusive collective excitations at THz frequencies.

**Materials and Methods**:

**Sample preparation.** Bare KTO and STO crystals, 500 μm thick, were obtained from MTI and Crystal GmbH, respectively, with KTO measuring 10 mm × 10 mm and STO 7.5 mm × 7.5 mm. In Py/KTO and Py/STO samples, a 3 nm Py layer was deposited on substrates at room temperature, using DC sputtering with 3 mTorr Ar pressure and a 0.7 Å/s rate. For Py/SiO$_2$/KTO, a 100 nm SiO$_2$ layer was RF-sputtered at 5 mTorr before depositing 3 nm Py. For Py/BTO/KTO, BTO layers of varying thicknesses were deposited using pulsed laser deposition at 700°C and an oxygen pressure of 15 mTorr. The laser energy density was maintained at 2.6 J/cm$^2$. After deposition, the samples were cooled to room temperature under the same pressure and transferred to the sputtering system. Then 3-nm-thick Py was deposited on the BTO layer using the same DC sputtering method. For Py/Al$_2$O$_3$/STO, amorphous Al$_2$O$_3$ films were deposited on STO substrates via atomic layer deposition (ALD) using trimethylaluminum and water precursors in a Veeco Fiji ALD tool at 150°C under rough vacuum (~0.2 Torr) with an Ar purge. A growth rate of 1.1 Å/cycle was confirmed by in situ spectroscopic ellipsometry on a Si wafer. Subsequently, 3 nm of Py was deposited on the Al$_2$O$_3$ layer using DC sputtering.

**THz generation and detection in SSTS.** A schematic of the THz setup and measurements is provided in Fig. S2. Samples were excited through the uncoated side by ~100 fs laser pulses with a wavelength of 800 nm (~1.55 eV) and a fluence of ~ 1.2 mJ/cm$^2$. The emitted THz field induced birefringence in a <110>-cut 1 mm-thick ZnTe crystal, resulting in a rotation of the probe beam's polarization. By employing a λ/4 waveplate, a Wollaston prism, and a pair of balanced photodiodes, the THz waves were detected using an electro-optic sampling technique. The THz setup was operated in a nitrogen gas-purged environment to ensure 0% relative humidity. The detected THz field $S(\omega)$ is a convolution of the emitted THz field $E_{near}(\omega)$ and the response function of the detection scheme $H(\omega)$: $S(\omega) = H(\omega)E_{near}(\omega)$. Here $H(\omega) = H_P(\omega)H_d(\omega)$, in which $H_P(\omega)$ accounts for the propagation of the THz pulse from the sample to the ZnTe crystal, $H_d(\omega)$ quantifies the electro-optic detection of the THz pulse by the probe beam in the ZnTe crystal. The response function $H(\omega)$ does not produce any pronounced spectroscopic peaks and dips within the THz detection window.

**Acknowledgments**

We thank discussions with Dr. Peter Littlewood, Dr. Gian G. Guzmán-Verri and Dr. Burak Guzelturk. **Funding**: This work (THz emission experimental design and execution, data collection, sample growth, theory development) was supported by the U.S. Department of Energy, Office of Science, Basic Energy Sciences, Materials Sciences and Engineering Division. C.Z., K.H., X.X., and H.W. acknowledge the support for Raman spectroscopy measurements and part of data analysis and manuscript preparation by the U.S. Department of Energy, Office of Science, Basic Energy Sciences, under contract number DE-SC-0012509. The use of electron microscopy by J.W. at the Center for Nanoscale Materials, U.S. Department of Energy, Office of Science User Facilities, was supported by the U.S. DOE, Office of Basic Energy Sciences, under Contract No. DE-AC02-06CH11357. **Author contributions**: H.W. conceived and supervised the project. Z.C. designed the experiments, built the THz spectroscopy setup, performed the THz measurements, and analyzed the data. J.Y., Y.L., and A. R. B. prepared the samples under the supervision of A.B., D.D.F., and A. M.. K.H. conducted Raman measurements under the supervision of X.X.. J.W. performed the cross-section transmission electron microscopy measurements. Q.Z. performed the preliminary THz measurements on STO samples. M.N. conducted the theoretical analysis. Z.C. wrote the paper with input from all authors, in particular, S.H., A.B., M.N., and H.W.

**Figure Captions**

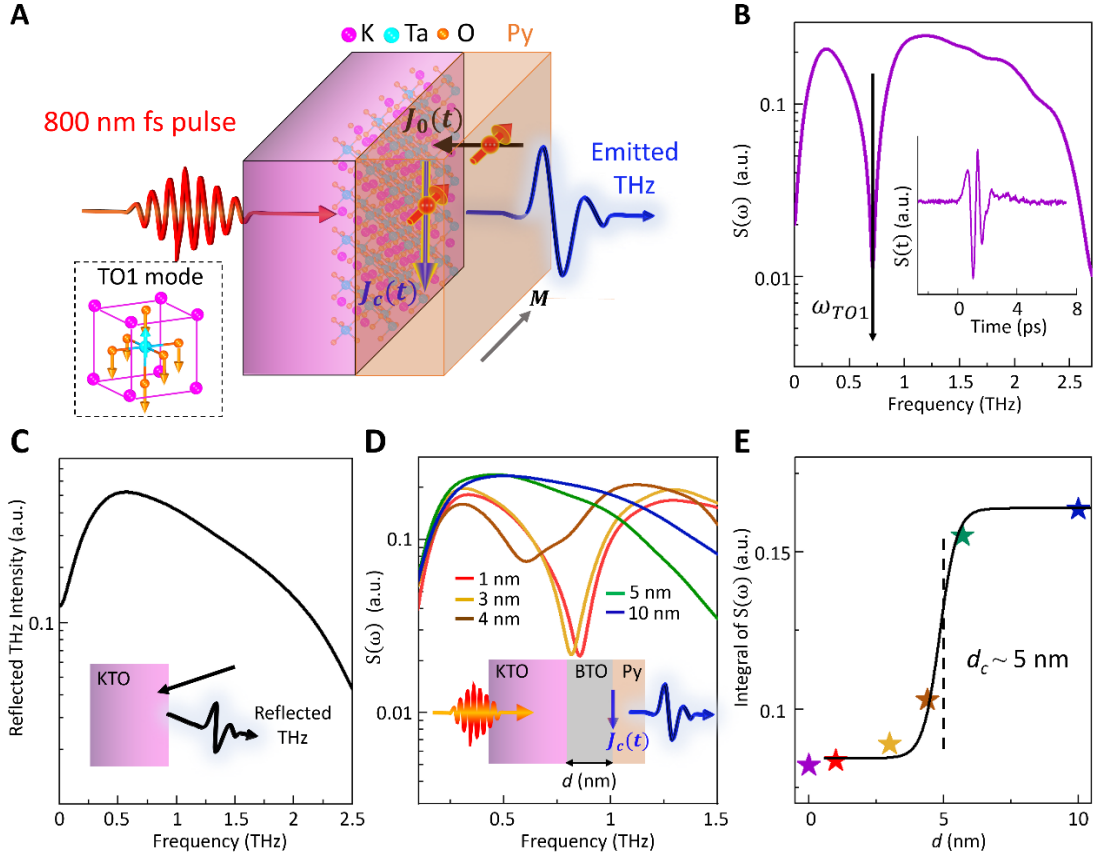

**Fig. 1. Surface-sensitive spintronic THz spectroscopy. A**, Schematic representation of the THz emission in SSTS, using a Py/KTO (111) sample as an example. Laser-excited spin-polarized hot electrons (i.e., $J_0(t)$) diffuse towards the Py-KTO interface and subsequently transform into a charge current $J_c(t)$ through spin-charge conversion. $J_c(t)$ is the source of the THz radiation. "M" and the gray arrow represent the in-plane magnetization of Py. The black dashed rectangle shows a diagram of the TO1 phonon mode in a KTO unit cell. **B**, THz emission spectrum from Py/KTO (111), where the dip indicates the TO1 mode frequency $\omega_{TO1}$. Inset: Corresponding THz time-domain waveform. **C,** Reflected THz spectrum at the KTO surface, where no TO1 phonon dip is observed. The inset shows a simple schematic of the THz reflection measurement (see Fig. S6 for more details). **D**, THz emission from Py/BTO-buffer/KTO (111) samples with various buffer thicknesses. The inset shows a schematic of the sample structures with a BTO-buffer layer between the KTO surface and the Py layer, where $d$ denotes the thickness of the buffer layer. **E**, Integral of the THz spectra versus the thickness of the spacer. A tanh function fitting $\int S(\omega)d\omega = a + b \tanh(\frac{d-d_c}{c})$ indicates that the probe depth $d_c$ of SSTS is about 5 nm, as marked by the dashed black line. The purple star in **E** presents the result with no buffer layer as shown in **B**. All the results shown in this Figure were taken at T = 10K, and the THz field was detected on the Py side.

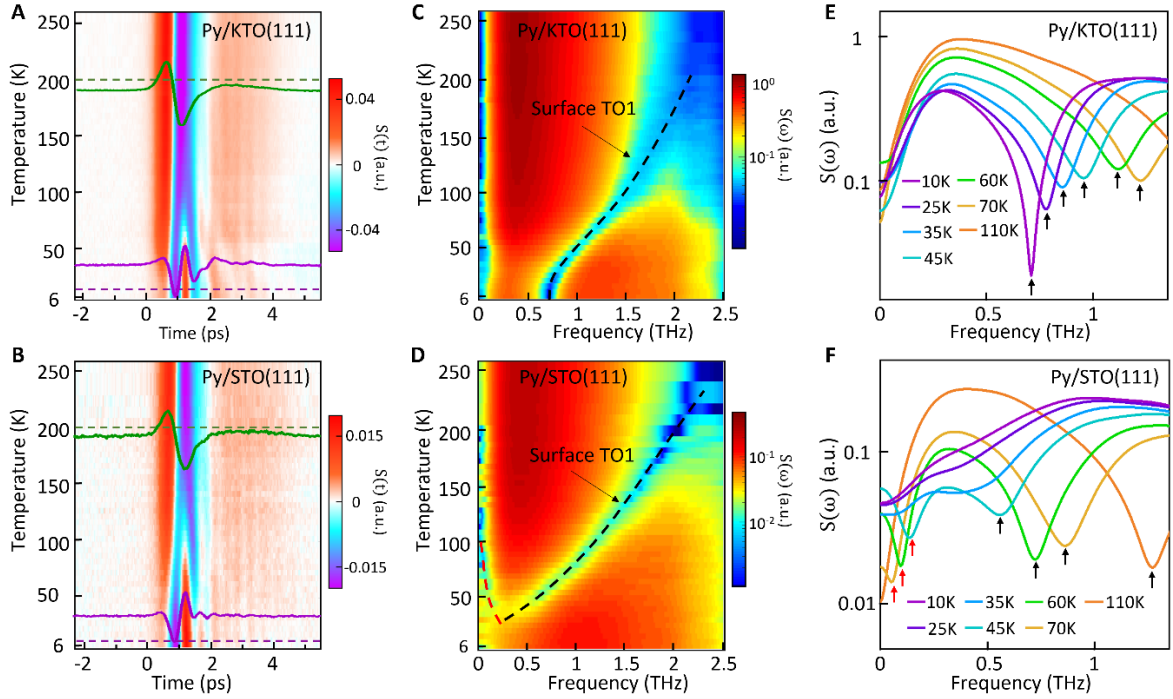

**Fig. 2. Visualization of sub-THz phonons at KTO (111) and STO (111) surfaces.** **A** and **B** display the images of the temperature-dependent THz waveforms of Py/KTO (111) and Py/STO (111) samples, respectively. **C** and **D** depict the Fourier spectra of **A** and **B**, respectively. The black dashed curves in **C** and **D** mark the softening of the surface TO1 mode of KTO and STO, respectively. The red dashed curve in **D** highlights the hardening of the mode related to antiferrodistortion at the STO surface. **E** and **F** represent line-cut THz spectral profiles at selected temperatures of KTO and STO samples, respectively. **E** illustrates the sharpening of the KTO surface TO1 mode spectral dip (marked by black arrows) as the temperature decreases, and **F** shows the broadening and shallowing of the STO surface TO1 mode spectral dip (marked by the black arrows) and the sub-meV mode spectral dip (marked by red arrows) at low temperatures.

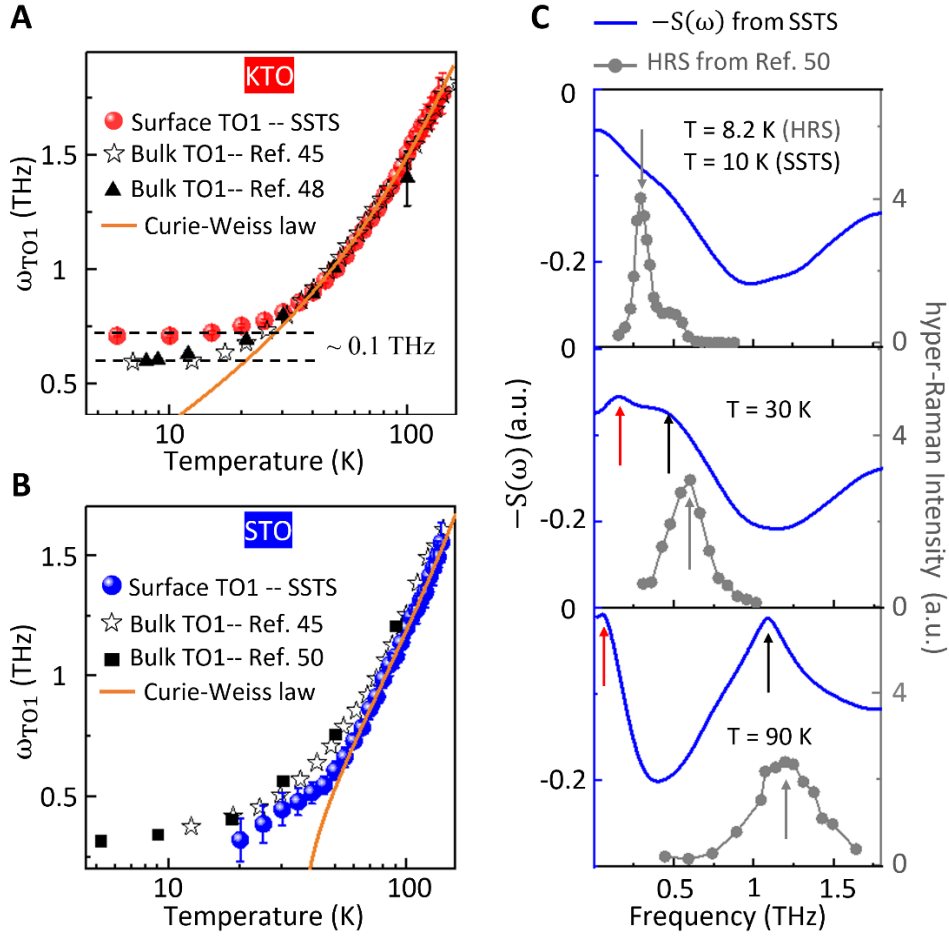

**Fig. 3. Temperature-dependent phonon frequencies in KTO and STO samples. A**, Comparison of the KTO surface TO1 and bulk TO1 mode frequencies. The data from THz transmission spectroscopy and HRS are adapted from Ref. 48 and Ref. 45, respectively. **B**, Comparison of STO surface TO1 and bulk TO1 mode frequencies. In SSTS measurements, the surface TO1 mode frequencies below 20 K could not be determined given the broadness of the spectral dip and its overlap with the sub-meV mode. The data of HRS are adapted from Refs. 45 and 50. The orange curves in **A** and **B** represent the Curie-Weiss law fitting $\omega_{TO1} \propto \sqrt{T - T_c}$ with $T_c \sim 5K$ in **A** and $\sim 37.5K$ in **B**, respectively. **C,** Comparison of spectral sharpness between surface TO1 and bulk TO1 modes in STO. The HRS measurements (adapted from Ref. 50) show a sharpening of the bulk TO1 mode in an STO crystal as the temperature decreases. The blue curves present plots of the reversed SSTS spectral data (i.e., $-S(\omega)$) of the Py/STO (111) sample from Fig. 2D, providing a direct comparison of spectroscopic peak-to-peak characteristics with Ref. 50. At lower temperatures, the SSTS spectra do not exhibit pronounced peaks, due to the broadening of the surface TO1 mode. Here the black (SSTS) and gray (HRS) arrows mark the TO1 mode positions, and the red (SSTS) arrows mark the surface sub-meV mode.

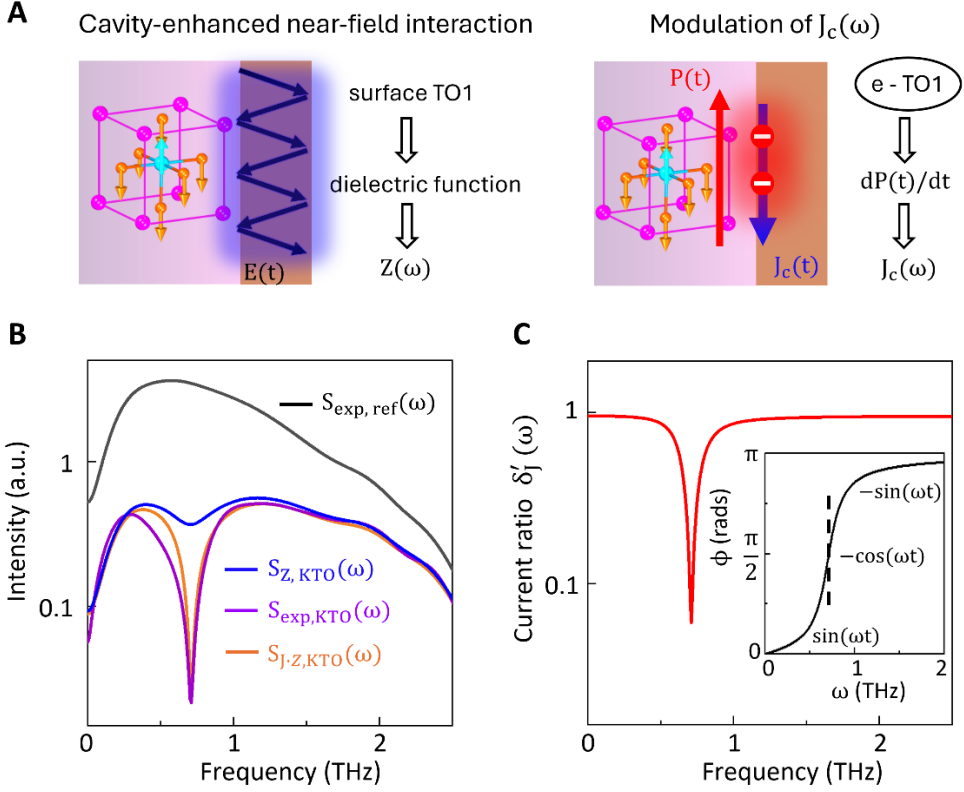

**Fig. 4. Theoretical Modeling. A**, Schematic of the detection mechanisms of SSTS. Left: Detection of surface modes through near-field interactions. The Py layer functions as a thin cavity where all reflected THz fields (black arrows) within the cavity constructively interfere, facilitating near-field THz-matter interaction. The blue glow of the THz field illustrates the depth of the near-field THz-matter interaction. The effective near-field THz within the cavity is represented by E(t) with its Fourier spectrum denoted as $E_{near}(\omega)$. $Z(\omega)$ is the cavity impedance, encapsulating the spectral information of the surface TO1 mode through the dielectric function. Right: Detection of surface modes via modulation of the surface current, $J_c(\omega)$. The spin-polarized electrons (the red balls) involved in the transient current $J_c(t)$ (the blue arrow) couple with the surface TO1 mode, so that $J_c(t)$ can induce a surface in-plane transient polarization P(t) with its derivative dP(t)/dt representing a transient current that destructively interferes with $J_c(t)$. The red glow indicates the tunneling depth of the hot electrons that determines the probing depth of $J_c(\omega)$. **B**, Comparison between the experimental data $S_{exp,KTO}(\omega)$ and $S_{exp,ref}(\omega)$ at 6 K, with calculated THz spectra: $S_{Z,KTO}(\omega)$ including the dielectric contribution alone; $S_{J\cdot Z,KTO}(\omega)$ including the contribution from both the dielectric response and the induced dP/dt term. **C,** Calculated $\delta'_J(\omega)$ profile with $c_{ind}$= 1.63×10⁻⁴, $\gamma_{TO1}$ = 0.24 THz used in **B**. The inset shows the phase angle (ϕ) of the dielectric function as a function of frequency ω. At $\omega_{TO1}$(the black dashed line), $\phi = \frac{\pi}{2}$, the induced current contribution ∝ −cos(ωt) and thus destructively interferes with the driving current $J_c(t) \propto$ cos(ωt).

# Supplementary Materials

**Fig. S1. Cross-sectional TEM images of a Py/KTO (111) sample**

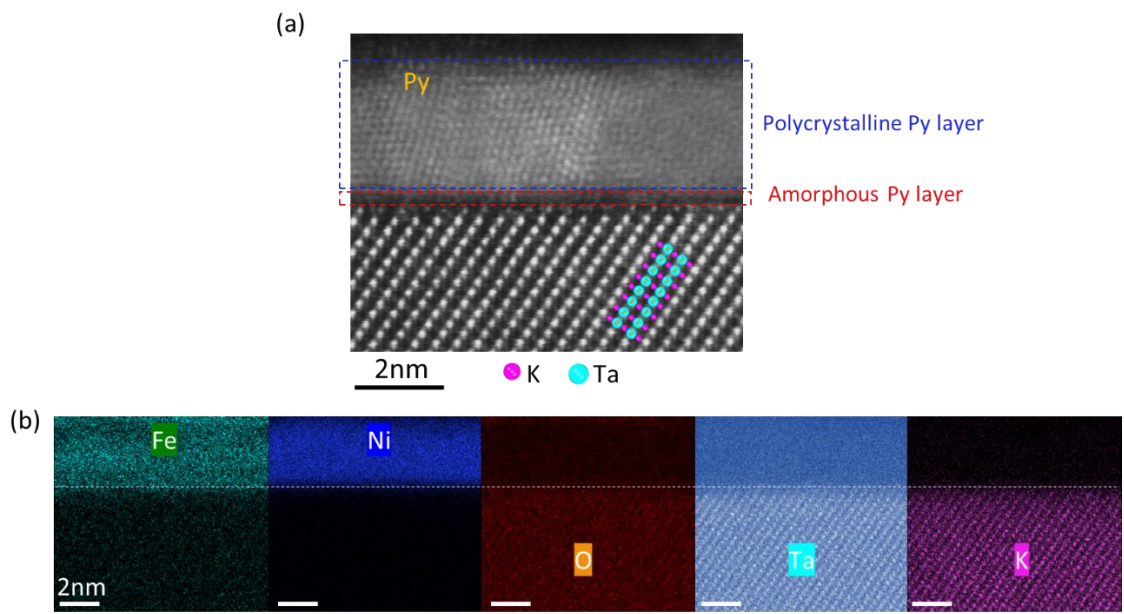

Fig. S1. (a) Cross-sectional high-angle annular dark-field (HAADF) scanning transmission electron microscopy (STEM) image of a Py/KTO (111) sample. (b) Elemental mapping images for the Py layer (Fe, Ni) and the KTO layer (K, Ta, O), illustrating the atomically precise interface between Py and KTO.

**Fig. S2. Experimental Setup**

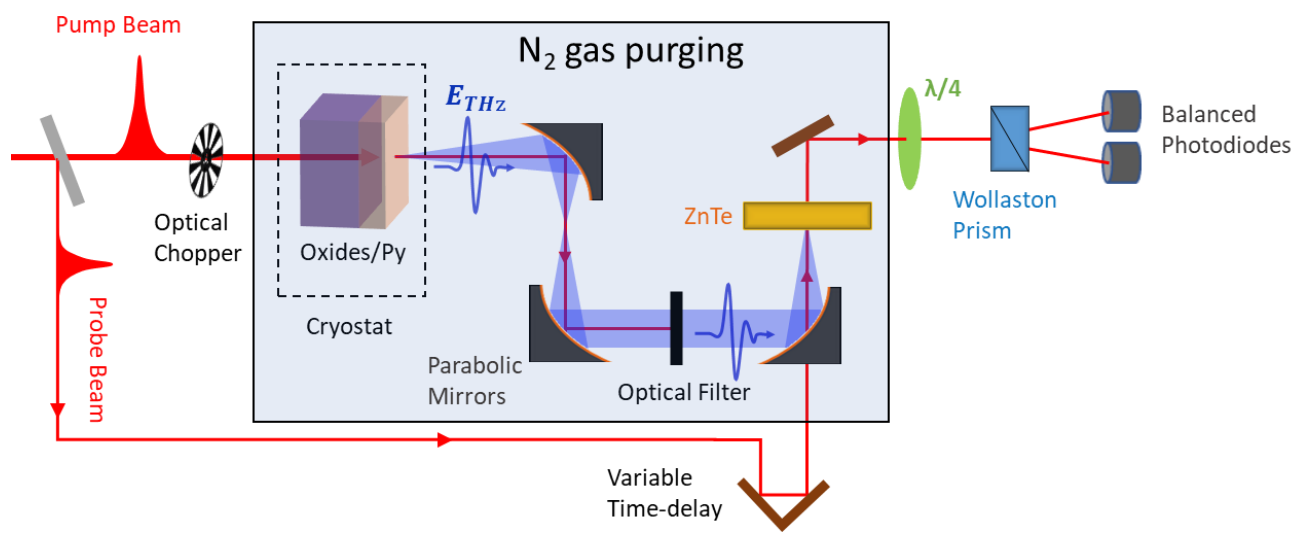

Fig. S2. A schematic of the THz spectroscopy setup.

**Fig. S3. Comparison of THz emission from Py/STO, bare STO, and Pt/STO samples**

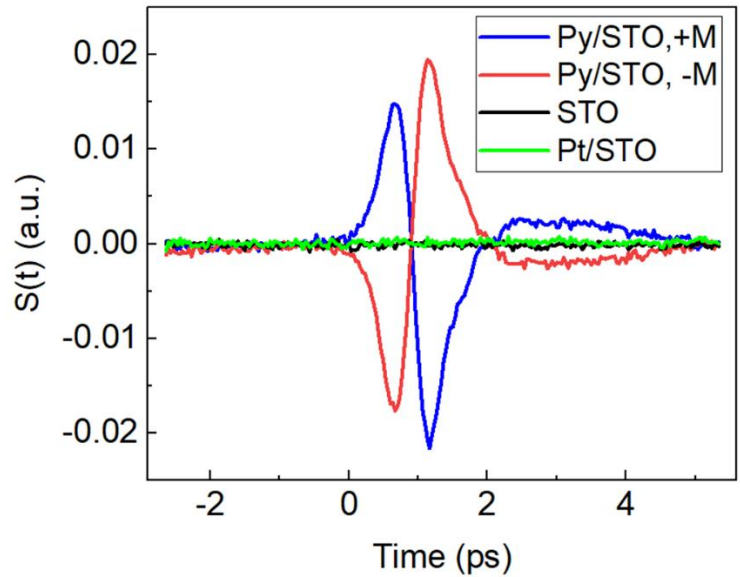

Fig. S3. Comparison of THz emission from Py/STO, bare STO, and Pt/STO samples. Pt: platinum. Here ±M labels the in-plane magnetization direction of the Py layer. Both Py and Pt are 3 nm thick. No THz signal is detected from the bare STO and Pt/STO samples, signifying the key role of spin-polarized electrons in the THz emission process.

**Fig. S4. Reversing the THz polarization by changing the magnetization M and turning the sample around, respectively**

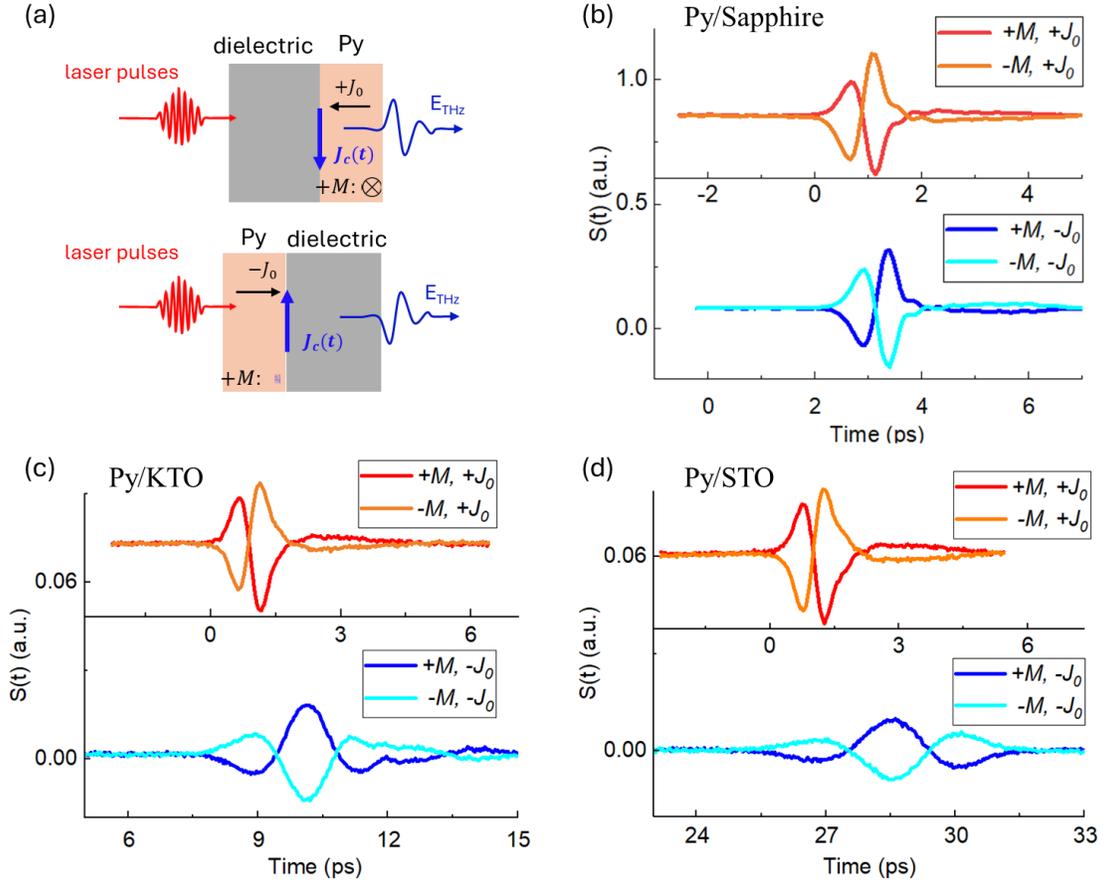

Fig. S4. (a) Schematic of THz emission from Py/dielectric samples under different pump-probe configurations. Top: Laser pumping from the dielectric side, with the spin diffusion current $J_0$ (black arrow) denoted as $+J_0$, and the magnetization $M$ direction (into the screen) denoted as $+M$. The interfacial transient current $J_c(t) \propto J_0 \times M$ is the source of THz radiation. Bottom: Laser pumping from the Py side by turning the sample $180^0$ about $M$, causing $J_0, J_c$ and the emitted THz to reverse their sign. The dielectric materials are sapphire (500 μm thick), KTO (200 μm thick) and STO (500 μm thick). (b-d) Time-domain THz waveforms at 295K, varying the magnetization directions (±M) and pump orientations (±$J_0$) in Py/sapphire (b), Py/KTO (c), and Py/STO (d). The reversal of the THz polarization in these samples upon rotating the sample (i.e., changing the sign of $J_0$) confirms that the contribution from ultrafast demagnetization-induced magnetic dipole radiation is negligible in the measured THz field. Instead, spin-charge conversion (i.e., $J_c(t) \propto J_0 \times M$) through mechanisms such as the inverse spin Hall effect, anomalous Hall effect, skew scattering, and inverse Edelstein effect at the interface, dominate the THz emission. The time delays of the THz pulses between $+J_0$ and $-J_0$ cases arise from the refractive index mismatch of 800nm and THz lights within the dielectric materials. The THz profiles in Py/sapphire are nearly identical, differing only in sign, due to the negligible THz absorption by sapphire. In contrast, KTO

and STO, with their larger refractive indices at THz frequencies, exhibit significant THz absorption in the "$-J_0$" configuration, leading to differences in the THz time traces compared to the "$+J_0$" configuration.

**Fig. S5. Temperature-dependent THz emission from a Py/sapphire sample**

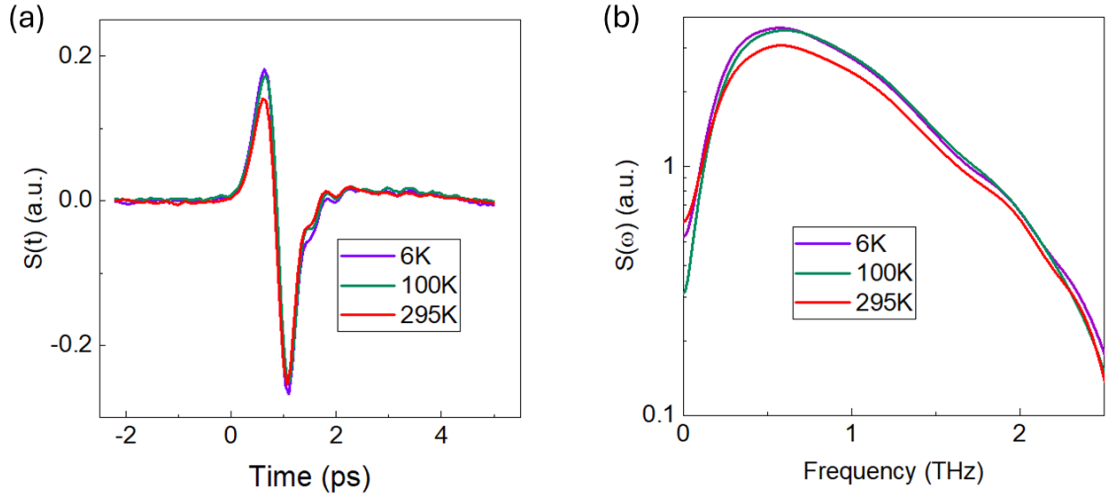

Fig. S5. Time-domain THz waveforms (a) and the frequency-domain THz intensity profiles (b) at 6K, 100K, and 295K. Minimal variations are observed between the THz signals at 6 K, 100K and 295 K, with no spectral dip features in the intensity profiles.

**Fig. S6. THz reflection measurements**

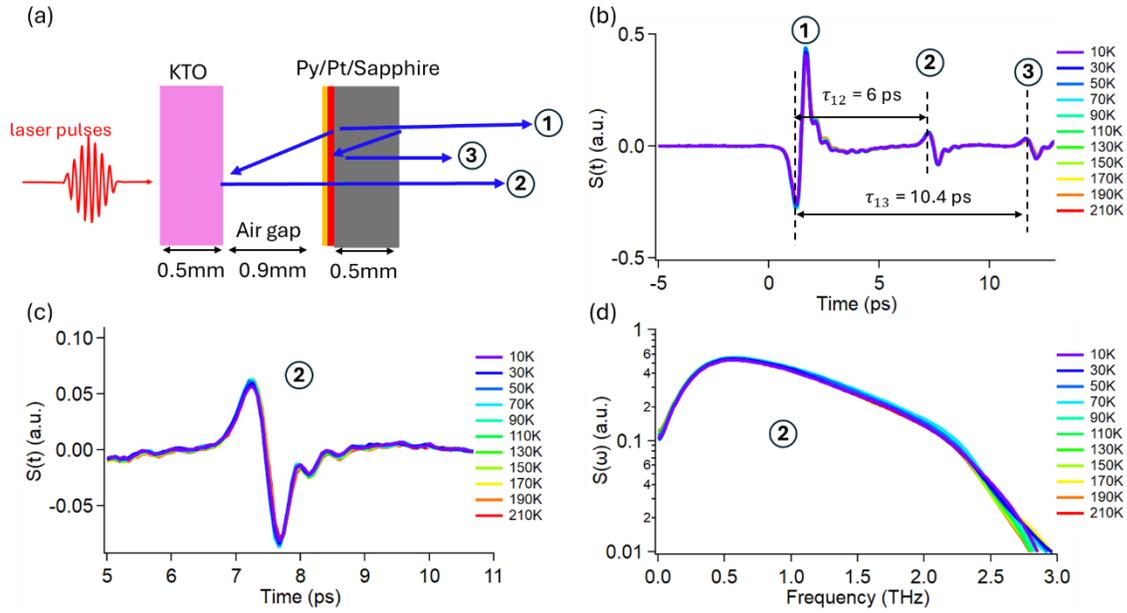

Fig. S6. Temperature-dependent THz reflection measurements. (a) Schematic of the THz reflection experiment setup. A Py/Pt/sapphire sample, commonly used as a spintronic THz emitter, serves as the THz source. It consists of a 3 nm thick Py layer, a 2 nm thick Pt layer, and a 0.5 mm thick ($d_{sap}$) sapphire substrate. The Py/Pt/sapphire sample is the THz source and placed close to a 0.5 mm thick KTO crystal with an air gap of $d_{air} = 0.9$ mm to separate the multi-reflected THz signals. The primary THz pulse emission is marked as ①, while ②, and ③ represent the subsequent reflected pulses from the interfaces/surfaces, as labeled in (a). Notably, ② is the reflection from the KTO surface. The time delays between these pulses and the primary pulse are: $\tau_{12} = n_{air} \cdot 2d_{air}/v_{THz} = 6$ ps, and $\tau_{13} = n_{sap} \cdot 2d_{sap}/v_{THz} = 10.4$ ps, where the refraction indexes of air and sapphire are $n_{air} = 1$ and $n_{sap} \sim 3$, respectively, $d_{air} = 0.9$ mm, $d_{sap} = 0.5$ mm, and $v_{THz} = 3 \times 10^8$ m/s. (b) Time-domain THz profiles for pulses ① to ③ at various temperatures, showing little temperature dependence of the THz signals. (c) a zoom-in plot of the reflected pulse ②. (d) THz intensity profiles of the reflected pulse ② in the frequency domain, showing no spectral dip. These results demonstrate that the observed TO1 spectral dip in Fig. 1B is not due to the far-field absorption of the reflected THz wave.

**Fig. S7: SSTS spectrum in the samples with THz-transparent spacers inserted**

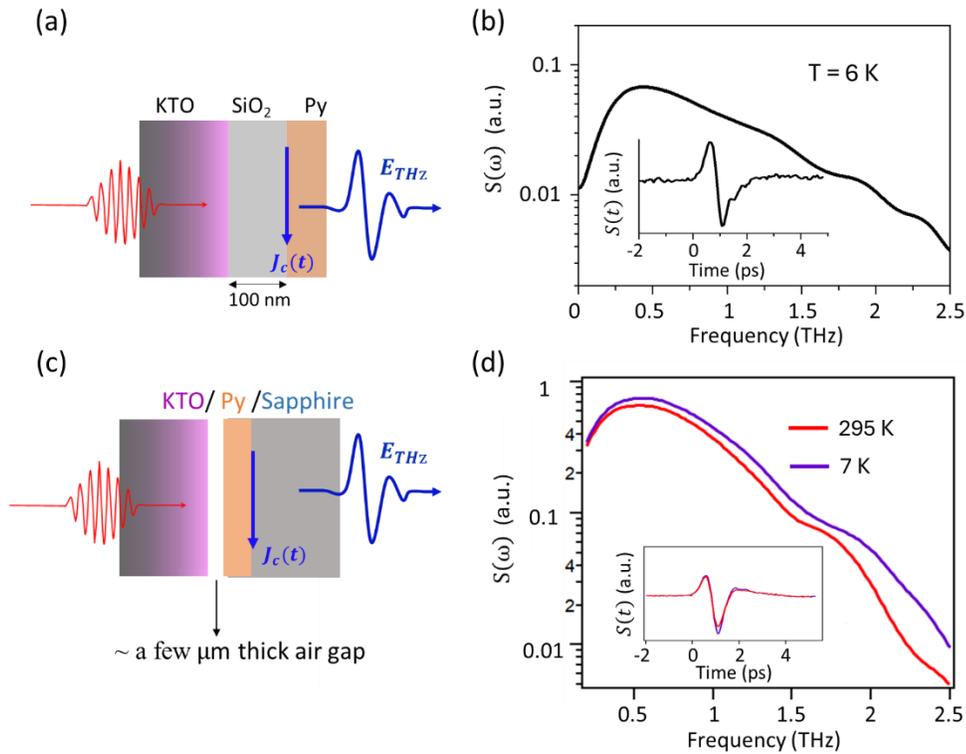

Fig. S7. THz emission from the Py/SiO$_2$/KTO (111) and KTO (111)/Py/sapphire samples. (a) shows a schematic of the SSTS measurements on the Py/100nm-thick-SiO$_2$/KTO (111) sample, where the THz field is detected at the Py side. (b) THz emission spectrum from the Py/100nm-SiO$_2$/KTO (111) sample at T = 6 K, where the TO1 mode is not detected. Inset: Time-domain waveform. (c) Schematic of the KTO (111) /Py/sapphire sample structure. a 3nm Py layer was deposited on one side of the sapphire substrate (500 μm thick). Subsequently, the KTO crystal was stacked on the Py-coated side, resulting in a KTO/Py/sapphire stacked configuration. This arrangement creates a few μm thick air gap between the Py and KTO surfaces. Such a gap separates the transient current J$_c$(t) and the KTO surface. We illuminated the sample from the KTO side and detected the emitted THz field on the sapphire side. The backward-emitting component of the THz field traversed the vacuum gap, reached the KTO surface, and was then reflected towards the Py layer. The detected THz waves at temperatures of 7 K and 295 K are presented in (d), showing the minimal variation in the THz signal between 7 K and 295 K, and the absence of dip features in the THz intensity profiles. Our results show that the THz transparent spacer layer (e.g., 100 nm thick SiO$_2$ and a few μm thick air gap) blocks the detection of TO1 mode.

**Fig. S8. Characterizations of Py/BTO/KTO samples**

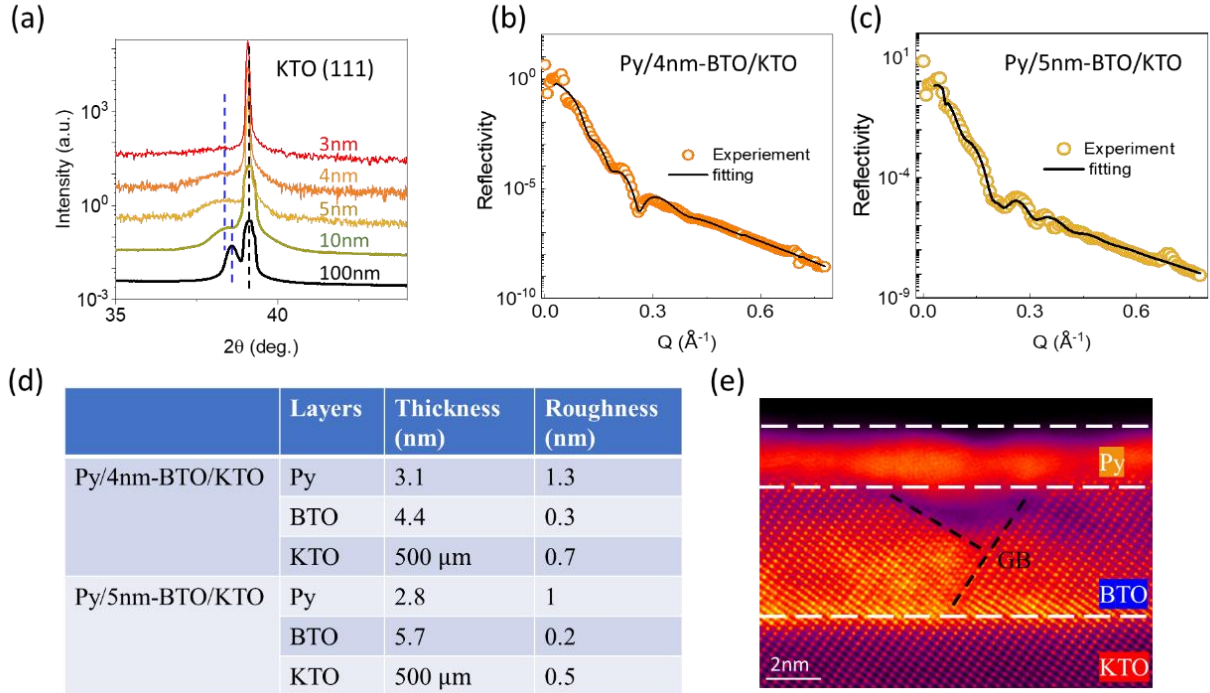

Fig. S8 (a) X-ray diffraction (XRD) data of Py/BTO/KTO (111) samples with varying BTO thicknesses (3 nm, 4 nm, 5 nm, 10 nm, and 100 nm). The X-ray energy is 8.04 keV. The 111 peak position of KTO is consistent with the measurements in Ref. 61. The 111 Bragg peaks of BTO, marked by blue dashed lines, are less prominent in thinner samples. The peak positions (2θ angle) for the 3-10 nm BTO films are ~ 0.2 degrees lower compared to the 100 nm BTO film, whose peak position is consistent with that of a non-strained BTO crystal[62,63]. (b) and (c) present the X-ray reflectivity (XRR) data for Py/4nm-BTO/KTO and Py/5nm-BTO/KTO samples, respectively. The solid lines represent the fitting curves, which were used to determine the thickness and roughness of each layer, as shown in (d). Note that the XRR intensity at low angles is slightly higher than 1 due to data correction with an error. (e) shows a cross-sectional HAADF STEM of the Py/4nm-BTO/KTO sample, where the grain boundaries (GB) in the BTO film are formed due to releasing the large lattice strain.

**Fig. S9. Details of the FFT procedure and determining the error bar of the measured TO1 mode frequency**

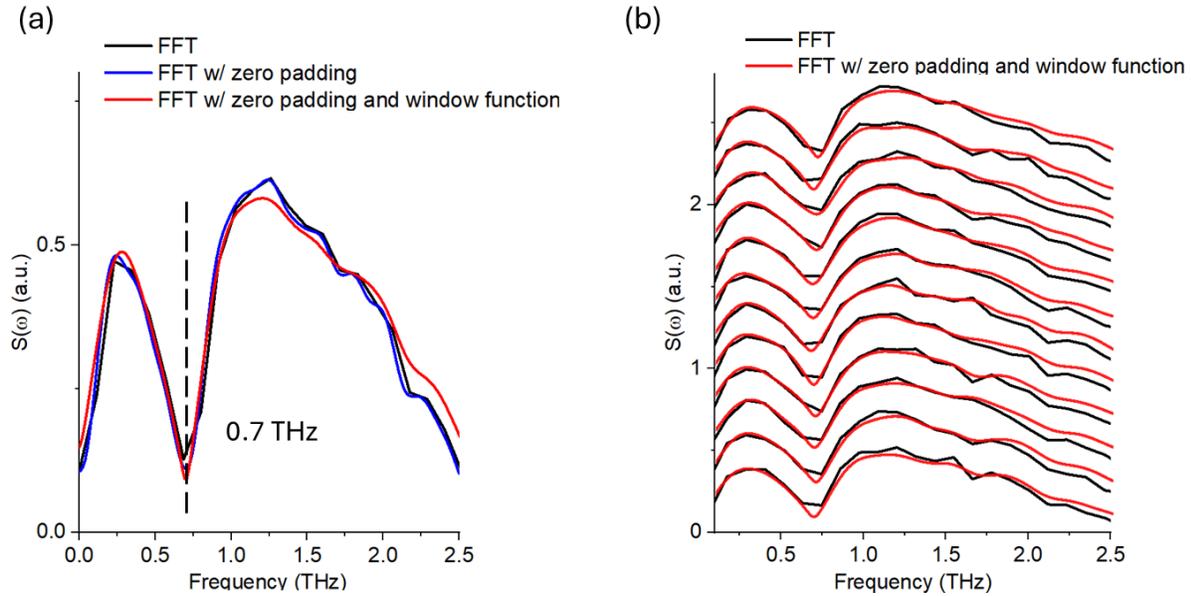

Fig. S9. Details of FFT procedure and estimation of the error bar for the measured TO1 mode frequency using data at 10K as an example. In this study, all THz time-domain trace data were averaged from more than ten individual repeated measurements to ensure consistency and a better signal-to-noise ratio. (a) Comparison of FFT results with and without zero padding and the Hanning window function. The black curve represents the FFT magnitude $S(\omega)$ of the averaged THz time-domain signal $S(t)$ from twelve repeated measurements, with a frequency step size of 0.11 THz. The blue curve shows the FFT with zero padding, which extends the time window for FFT, providing finer frequency resolution (~ 0.02 THz) without introducing artifacts. The red curve illustrates the FFT with both zero padding and the Hanning window function applied. These three FFT procedures yield very similar THz spectra in terms of shape and dip strength. For improving frequency resolution, all the frequency domain THz data presented in the main text and other figures in the Supplementary Materials were produced using both zero padding and the Hanning window function. (b) The THz spectra of the twelve individual repeated measurements. The black curves are the FFT magnitudes without zero padding and the Hanning window function. The red curves are the FFT with zero padding and the Hanning window function applied, from which, the TO1 mode frequency was determined as $0.7 \pm 0.02$ THz within a 99% confidence interval. The above analysis shows that we can determine the mode frequency with a higher frequency resolution in the FFT spectrum than the frequency interval determined by the length of the time delay window measured in the experiment.

**Fig. S10: A complete set of temperature-dependent THz emission spectrum from a third Py/KTO (111) sample and a Py/STO(001) sample**

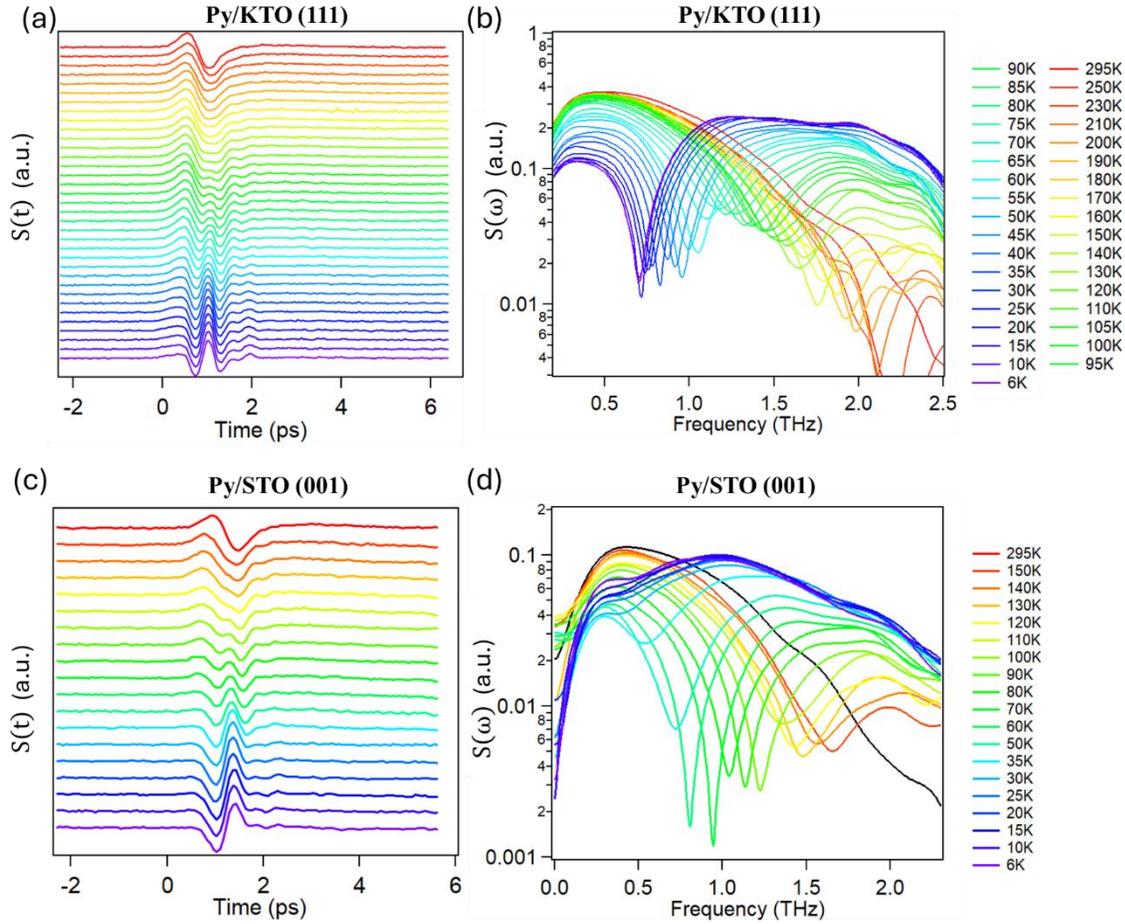

Fig. S10. Temperature-dependent THz field of another Py/KTO (111) sample represented in the time domain (a) and frequency domain (b). The thickness of this KTO crystal is 200 μm. The temperature-dependent surface TO1 mode in this sample agrees with that presented in the main text, where the KTO is 500 μm in thickness. The mode frequency levels off at ~ 0.7 THz when the temperature is below 15K. Temperature-dependent THz field from a Py/STO (001) sample depicted in the time domain (c) and frequency domain (d). The dips, corresponding to the surface TO1 mode, become broader, less pronounced, and imperceptible at temperatures below 35 K. Note that the spectral dip feature of the sub-meV mode is not obvious in this Py/STO (001) sample. The orientation dependence of this sub-meV mode needs further investigation, but is not unexpected given the known sensitivity of the antiferrodistortive transition in STO to surface orientation.

**Fig. S11: Raman measurements of a Py/STO (001) sample**

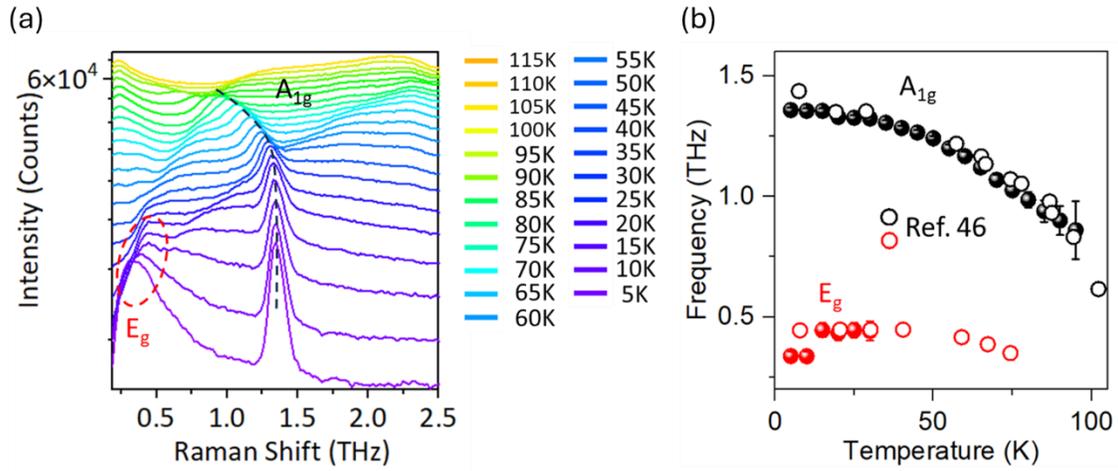

Fig. S11. (a) Temperature-dependent Raman scattering spectrum of a Py/STO (001) sample. Below the cubic to tetragonal phase transition temperature of 105 K in bulk STO, the soft R point phonon associated with it becomes mirrored at the $\Gamma$-point due to zone folding and splits into two Raman-active components of $A_{1g}$ (marked by the black dashed curve) and $E_g$ (highlighted by the red dashed circle) symmetries. (b) The temperature-dependent frequencies of $A_{1g}$ and $E_g$ in the Py/STO (001) sample, aligning with previous Raman measurements taken for bulk STO[46].

**Fig. S12: THz emission spectra from Py/AlO$_x$/KTO (001) and Py/AlO$_x$/KTO (111) samples**

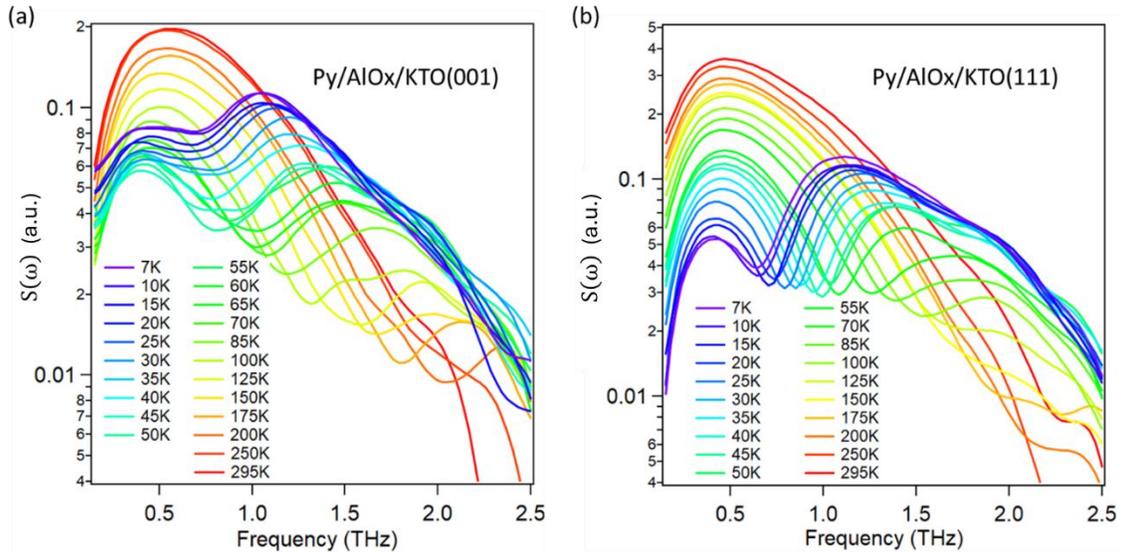

Fig. S12. Comparison of THz emission spectra from KTO 2DEG samples. The 2DEG is formed at the AlO$_x$(2 nm)/KTO interface due to oxygen vacancies. The 2DEG at the AlO$_x$/KTO (001) interface remains in the normal state down to 25 mK, and the 2DEG at the AlO$_x$/KTO (111) interface[3,4] becomes superconducting below ~ 2 K. Interestingly, at low temperatures, the spectral dips from the KTO TO1 mode at the (001) surface are much broader and shallower than those at the (111) surface.

**Fig. S13. The contribution from the dielectric response in SSTS**

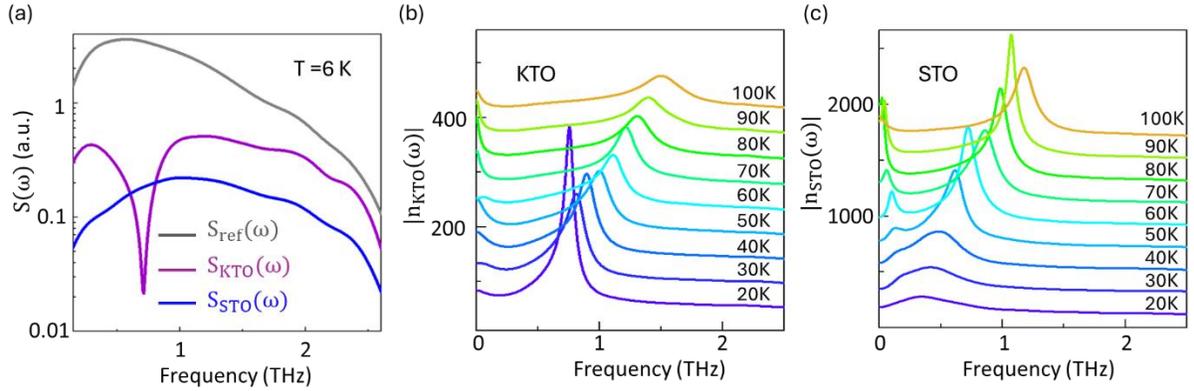

Fig. S13. (a) Comparison of the THz spectral intensity $S(\omega)$ of Py/KTO (purple) and Py/STO (blue) to a Py/sapphire reference (gray). The THz signal from KTO and STO is weaker because the refraction indices of KTO and STO are larger than that of sapphire. (b) and (c) show the calculated modulus of the refractive index of KTO surface ($n_{KTO}(\omega)$) and STO surface ($n_{STO}(\omega)$), respectively, by considering the contribution of the dielectric response alone. We estimate $Z_0 G = 1.7$ based on extrapolating the conductance of Py films to the thickness of 3 nm[64] and then exploit Eq. (2) to estimate $|n_{samp}(\omega)|$ (formally $|n_{samp}+Z_0G+1|$). These values assumed C=1 in Eq.(2) and would need to be multiplied by the current ratio (estimated to be C=1.62 from Fig. 4) for a better estimation. (b) illustrates the peak in the refraction index of KTO due to the TO1 mode that grows in strength and softens in energy as the temperature is lowered. (c) shows the refraction index of STO as a function of temperature. Similar to KTO, a TO1 mode is present and softens as the temperature decreases. But below about 100 K, a low-energy mode near 0.1 THz shows up that hardens as T decreases, while both this mode and the TO1 phonon broaden. Around 35 K, these two peaks merge into a single broad hump. This is consistent with the absence of the spectral dip at low T in Figs. 2D and 2F and implies an interaction between these two modes that is not evident in the bulk.

**Fig. S14. THz emission with $Al_2O_3$ buffer layers**

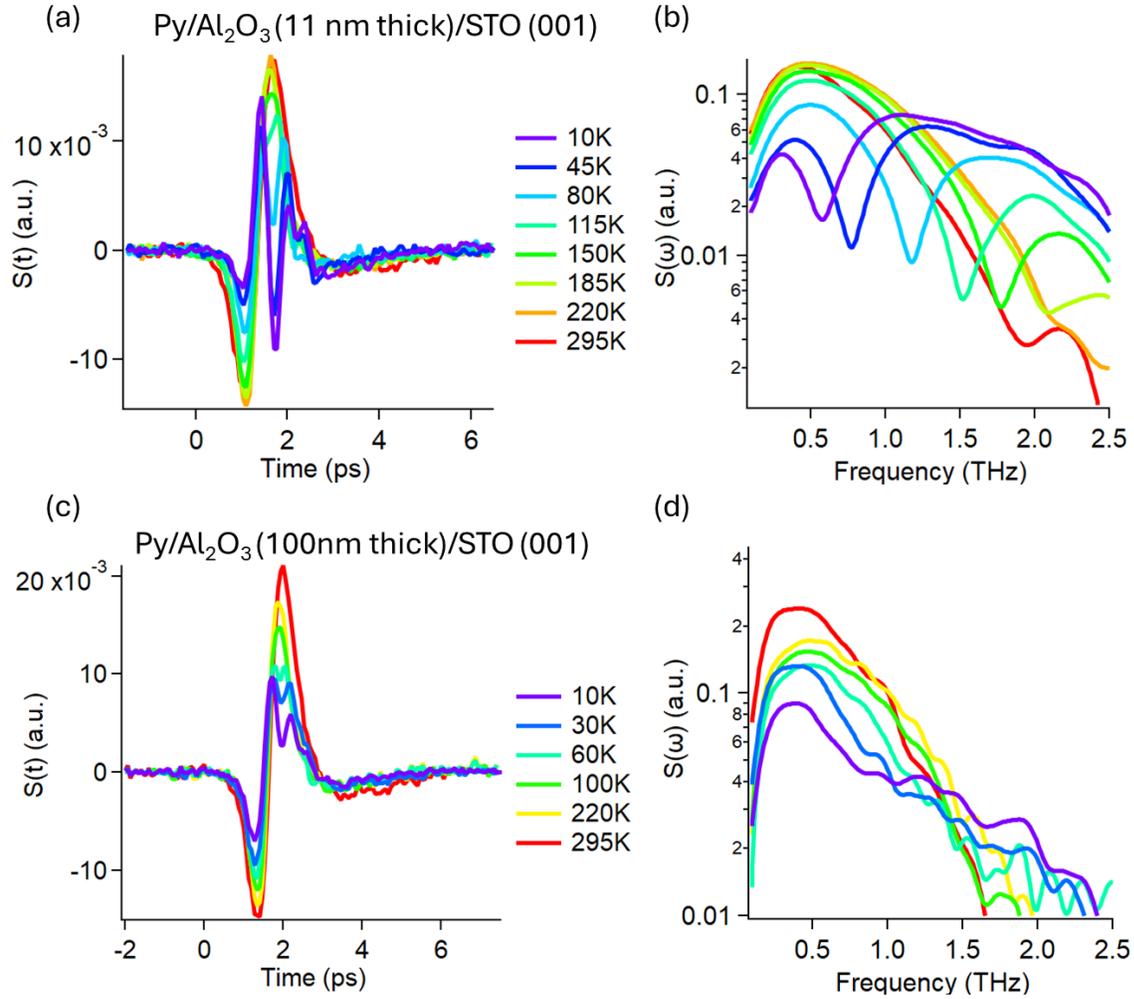

Fig. S14. THz emission in samples with a $Al_2O_3$ buffer layer. (a) and (b) THz emission spectrum of a Py/$Al_2O_3$ (11 nm thick)/STO (001) sample. The $Al_2O_3$ layer was deposited using atomic layer deposition (ALD). The transient current $J_c(t)$ is generated at the Py/$Al_2O_3$ interface, as the 11 nm-thick $Al_2O_3$ layer effectively blocks spin-polarized electron tunneling from the Py layer to the STO surface. The presence of the TO1 mode of STO should thus be attributed solely to the dielectric response, i.e. $Z(\omega)$. The dielectric constant of $Al_2O_3$ is much smaller than that of BTO[40], allowing the probe depth of near-field THz-matter interaction in $Al_2O_3$ to be deeper than that in BTO (Fig. 1D). (c) and (d) THz emission spectrum of a Py/$Al_2O_3$ (100 nm thick)/STO (001) sample. The $Al_2O_3$ layer in this sample was also deposited with ALD. Although slight changes in the time trace profiles were observed as the temperature decreased, the spectral dip is not prominent in the frequency domain. This suggests that a 100 nm $Al_2O_3$ layer deposited via ALD can block the detection of the TO1 mode for both $Z(\omega)$ and $J_c(\omega)$.